\documentclass[prl,aps,twocolumn,showpacs,floatfix,superscriptaddress]{revtex4}
\usepackage{graphicx}
\usepackage{amsmath}
\usepackage{amssymb}
\usepackage{color}

\begin{document}

\title{Decay of fermionic quasiparticles in one-dimensional quantum
  liquids}

\author{K. A. Matveev} 

\affiliation{Materials Science Division, Argonne National Laboratory,
  Argonne, IL 60439, USA}

\author{A. Furusaki}

\affiliation{Condensed Matter Theory Laboratory, RIKEN, Wako, Saitama
  351-0198, Japan}
\affiliation{RIKEN Center for Emergent Matter Science (CEMS), Wako,
  Saitama 351-0198, Japan}

\date{December 19, 2013}

\begin{abstract}

  The low-energy properties of one-dimensional quantum liquids are
  commonly described in terms of the Tomonaga-Luttinger liquid theory,
  in which the elementary excitations are free bosons.  To this
  approximation the theory can be alternatively recast in terms of
  free fermions.  In both approaches, small perturbations give rise to
  finite lifetimes of excitations.  We evaluate the decay rate of
  fermionic excitations and show that it scales as the eighth power of
  energy, in contrast to the much faster decay of bosonic excitations.
  Our results can be tested experimentally by measuring the broadening
  of power-law features in the density structure factor or spectral
  functions.

\end{abstract}


\maketitle

According to Landau's Fermi liquid theory \cite{landau1956Fermiliquid,
  landauStat_Mech_II} the low-energy properties of systems of
interacting fermions can be described in terms of a gas of weakly
interacting quasiparticles.  The latter retain fermionic statistics,
and their energy depends linearly on the momentum $p$ measured from
the Fermi surface, $\varepsilon=v(p-p_F)$.  A quasiparticle can decay
by exciting a particle-hole pair, but the resulting decay rate
$\tau^{-1}\propto\varepsilon^2$ is small compared to the energy.

For a one-dimensional system with linear spectrum, conservation of
momentum automatically implies conservation of energy, leading to a
divergent decay rate.  This results in a failure of the Fermi liquid
theory in one dimension.  Instead, one-dimensional systems are
commonly treated in the framework of the Tomonaga-Luttinger liquid
theory \cite{haldane1981luttinger, giamarchi2004quantum}, where the
elementary excitations are bosons.  In terms of original fermions, the
bosons correspond to the particle-hole pairs.  At small momentum $q$,
excitation energy is a linear function of $q$, and the Hamiltonian of
the system is given by
\begin{equation}
  \label{eq:Luttinger_Hamiltonian}
  H_{TL}=\sum_q  v|q|b_q^\dagger b_q^{},
\end{equation}
where $b_q$ is the boson annihilation operator.  In the case of a
system of interacting one-dimensional fermions with linear spectrum,
the Hamiltonian (\ref{eq:Luttinger_Hamiltonian}) can be derived using
the standard bosonization procedure \cite{haldane1981luttinger,
  giamarchi2004quantum}.

In general, however, the spectrum of the original particles is not
linear, and in addition to $H_{TL}$ the full Hamiltonian contains
corrections that account for the effect of the curvature of the
spectrum.  At low energies, these corrections are small and can often
be neglected.  For example, the simplest correction contains terms
cubic in bosons, $b_{q_1+q_2}^\dagger b_{q_1}b_{q_2}$, and represents
an irrelevant perturbation to the Hamiltonian
(\ref{eq:Luttinger_Hamiltonian}).  On the other hand, this and other
irrelevant perturbations enable the decay of bosonic excitations.
Thus, similarly to the quasiparticles in a Fermi liquid, bosonic
excitations in a Luttinger liquid have a finite decay rate.  The
evaluation of the lifetimes of bosonic excitations is a challenging
problem.  However, the basic result for the boson decay rate
$\tau^{-1}\propto\varepsilon^2$ \cite{samokhin_lifetime_1998} can be
understood simply as an uncertainty $q^2/m$ of the energy of the
particle-hole pair with momentum $q$ caused by the curvature of the
spectrum near the Fermi point.

Instead of describing the properties of a Luttinger liquid in terms of
bosonic excitations, one can formulate an alternative theory based on
quasiparticles with Fermi statistics.  This is accomplished by
noticing that the Hamiltonian (\ref{eq:Luttinger_Hamiltonian}) gives
an exact description of the excitations of a system of noninteracting
fermions with linear dispersion \cite{mattis_exact_1965,
  rozhkov_fermionic_2005}.  The new fermionic excitations coincide
with the original particles if the latter do not interact.  
It is important, however, that in systems with arbitrarily strong
interactions, the quasiparticles are only weakly interacting, in
analogy with the Fermi liquid theory in higher dimensions.  The
interactions result in scattering of the fermionic excitations and
give rise to a finite lifetime.

In this paper we evaluate the decay rate of the fermionic excitations
in a one-dimensional quantum liquid and show that it scales with
energy as $\tau^{-1}\propto\varepsilon^8$.  At low energies the
respective lifetimes are much longer than those of bosonic
excitations.  We therefore show that the fermionic picture has a
significant advantage over the conventional bosonic one when the
curvature of the spectrum is important.  Experimentally, the decay of
fermionic excitations should manifest itself as broadening of sharp
features at the quasiparticle mass shell in the density structure
factor and spectral functions.

The various phenomena caused by spectral curvature in one-dimensional
systems have been the subject of active study in the last few years
\cite{imambekov_one-dimensional_2012}.  Notably, the decay rate of
excitations in a weakly interacting Fermi gas was studied in
Ref.~\onlinecite{khodas2007fermi}.  In a system with quadratic
correction to the spectrum of the fermions, the decay of hole
excitations at zero temperature is forbidden by the conservation laws,
whereas for the particle excitations the result
$\tau^{-1}\propto\varepsilon^8$ was obtained.  Unlike
Ref.~\onlinecite{khodas2007fermi}, we are interested in a system with
arbitrary interaction strength.  It is important to stress that in
this case the fermionic excitations are not the original particles
studied in Ref.~\onlinecite{khodas2007fermi}, but the true
quasiparticle excitations of the Luttinger liquid defined via the
refermionization procedure \cite{rozhkov_fermionic_2005}.

The simplest way to introduce fermionic quasiparticles in the
Tomonaga-Luttinger liquid is by noticing that its Hamiltonian
(\ref{eq:Luttinger_Hamiltonian}) has the same basic form for both
interacting and noninteracting fermions, although the formal
transformation from fermions to bosons does depend on interactions.
Let us consider a system of free fermions described by the Hamiltonian
\begin{equation}
  \label{eq:H_0}
  H_0=\sum_k\varepsilon_k a_k^\dagger a_k^{}
      +\sum_p\tilde\varepsilon_{p} \tilde a_p^\dagger \tilde a_p^{},
\end{equation}
Here $a_k$ and $\tilde a_p$ are the fermion operators for particles on
the right- and left-moving branches, with momenta $k$ and $p$ measured
from the respective Fermi points.  If the spectra of the fermions are
linear, $\varepsilon_k=vk$ and $\tilde\varepsilon_p=-vp$, the
Hamiltonian (\ref{eq:H_0}) can be brought to the form
(\ref{eq:Luttinger_Hamiltonian}) using the standard bosonization
prescription \cite{haldane1981luttinger, giamarchi2004quantum}
\begin{subequations}
  \label{eq:bosonization}
\begin{eqnarray}
  b_q&=&i\left(\frac{2\pi\hbar}{qL}\right)^{1/2}
        \sum_k a_k^\dagger a_{k+q}^{},
\quad
  q>0,
\\
  b_q&=&-i\left(\frac{2\pi\hbar}{|q|L}\right)^{1/2}
        \sum_p \tilde a_{p}^\dagger \tilde a_{p+q}^{},
\quad
  q<0,
\end{eqnarray}
\end{subequations}
where $L$ is the system size.  For nonvanishing interactions, the
bosonization transformation is somewhat more complicated.  It is
controlled by the so-called Luttinger liquid parameter $K$, which
takes values $K<1$ for repulsive interactions and $K>1$ for the
attractive ones \cite{giamarchi2004quantum}.

Since the bosonic Hamiltonian (\ref{eq:Luttinger_Hamiltonian}) is
equivalent to the fermionic Hamiltonian (\ref{eq:H_0}) with linear
spectrum, instead of the standard Tomonaga-Luttinger liquid theory
with bosonic excitations $b_q$ one can construct an equivalent theory
based on the free fermion quasiparticles $a_k$ and $\tilde a_p$.  The
latter will coincide with the original fermions constituting the
Luttinger liquid only in the absence of interactions, i.e., at $K=1$.

In most physical systems the spectrum is not strictly linear.  To
account for the curvature, one has to add to the Hamiltonian
(\ref{eq:Luttinger_Hamiltonian}) various perturbations that are
irrelevant in the renormalization group sense.  Such perturbations are
given by operators with scaling dimensions higher than 2.  Because
each bosonic operator is mapped to a pair of fermion operators on the
same branch [see Eq.~(\ref{eq:bosonization})], the irrelevant
perturbations have even numbers of fermion operators on each branch.
In the simplest case, the perturbation consists of two fermion
operators, e.g., $k^la_k^\dagger a_k^{}$ with $l=2,3,\ldots$.  These
perturbations can be included into the Hamiltonian (\ref{eq:H_0}) by
allowing for the nonlinear quasiparticle spectrum
\begin{equation}
  \label{eq:varepsilon}
  \varepsilon_k=vk+\frac{k^2}{2m^*}+\frac{\lambda}{6}\,k^3+\ldots
\end{equation}
and $\tilde\varepsilon_p=\varepsilon_{-p}$.  

More complicated perturbations contain four, six, or more
quasiparticle operators.  For instance, the only such operator of
scaling dimension 3 is \cite{rozhkov_fermionic_2005}
\begin{equation}
  \label{eq:V_3}
  V_3=\frac{\gamma_3}{L}\sum_{kk'pp'}
      (k+k'-p-p')\,\delta_{k+p,k'+p'}
      \tilde a_{p'}^\dagger \tilde a_p^{} a_{k'}^\dagger a_k^{}.
\end{equation}
Here $\gamma_3$ is a constant, and the Kronecker delta reflects the
conservation of momentum. The perturbation (\ref{eq:V_3}) describes an
effective interaction of two quasiparticles.  Perturbations of higher
dimensions include interactions of any number of quasiparticles.
Because the interactions between the fermionic quasiparticles are
described by irrelevant perturbations, they are weak at low energies.

\begin{figure}
\includegraphics[width=.45\textwidth]{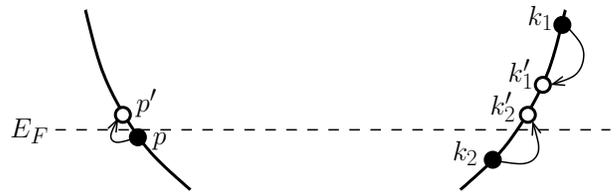}
\caption{A three-particle process leading to the decay of a
  quasiparticle in the state $k_1$ above the Fermi level involves two
  additional quasiparticles below the Fermi level, one, $k_2$, on the
  same branch and the other, $p$, on the opposite branch.}
\label{fig:decay-process}
\end{figure}

In the presence of interactions, the fermionic quasiparticles must
have a finite decay rate, which is the main subject of this paper.
Specifically, we consider a state with a filled Fermi surface of
states with $k<0$ and $p>0$ on the right- and left-moving branches,
respectively, and one additional quasiparticle in state $k_1>0$ on the
right-moving branch.  The two-particle scattering processes do not
simultaneously conserve the momentum and energy of the system, so the
simplest allowed process is three-particle scattering.  Furthermore,
at zero temperature two of the three quasiparticles must be on the same
branch, while the third is on the other branch \cite{khodas2007fermi},
Fig.~\ref{fig:decay-process}.

The decay rate is then found from Fermi's golden rule
\begin{eqnarray}
  \label{eq:golden_rule}
  \frac1\tau&=&\frac{\pi}{\hbar}
           \sum_{p,k_1',k_2'>0\atop p',k_2<0}
           \Big|{\mathcal A}_{k_1,k_2,p}^{k_1',k_2',p'}\Big|^2
\nonumber\\
          &&\times
           \delta(\varepsilon_{k_1}+\varepsilon_{k_2}+\tilde\varepsilon_{p}
                  -\varepsilon_{k_1'}-\varepsilon_{k_2'}-\tilde\varepsilon_{p'}),
\end{eqnarray}
where the transition matrix element is defined in terms of the $T$-matrix as
\begin{equation}
  \label{eq:matrix_element}
  {\mathcal A}_{k_1,k_2,p}^{k_1',k_2',p'}=
   \langle
    a_{k_1'} a_{k_2'} \tilde a_{p'}|T|\tilde a_{p}^\dagger a_{k_2}^\dagger a_{k_1}^\dagger
   \rangle.
\end{equation}
A three-particle scattering event can be accomplished in the second
order in perturbations coupling two particles, such as the term
(\ref{eq:V_3}).  In addition, a contribution to the amplitude
(\ref{eq:matrix_element}) can be obtained in the first order in
three-particle coupling, which appears in perturbations with scaling
dimensions higher than 4.

Because the conservation of momentum imposes restrictions on the final
states of the three particles, the matrix element
(\ref{eq:matrix_element}) must have the form $\mathcal
A\delta_{k_1+k_2+p,k_1'+k_2'+p'}$.  Furthermore, since we are
interested in the decay rate of a quasiparticle with small momentum
$k_1$, and all the other quasiparticles have momenta smaller than
$k_1$ in absolute value, one may expect to be able to take the limit
$k_1, k_1', \ldots\to0$ and replace $\mathcal A$ with the resulting
constant.  This would be incorrect because the quasiparticles on the
same branch are identical fermions, and thus the matrix element
(\ref{eq:matrix_element}) is antisymmetric with respect to
permutations $k_1\leftrightarrow k_2$ and $k_1'\leftrightarrow k_2'$.  
We therefore conclude that
\begin{equation}
  \label{eq:matrix_element_ansatz}
  {\mathcal A}_{k_1,k_2,p}^{k_1',k_2',p'}=
  \frac{\Lambda}{L^2}(k_1-k_2)(k_1'-k_2')\delta_{k_1+k_2+p,k_1'+k_2'+p'},
\end{equation}
where $\Lambda$ is symmetric with respect to the above permutation and
takes a constant value at $k_1\to0$.  We note that in the limit of
weak interactions the three-particle scattering amplitude does have
the form (\ref{eq:matrix_element_ansatz}) at $k_1\to0$, cf. Eq. (47) in
Ref.~\onlinecite{khodas2007fermi}.

Substituting the scattering amplitude (\ref{eq:matrix_element_ansatz})
into Eq.~(\ref{eq:golden_rule}) we obtain the expression
\begin{equation}
  \label{eq:decay_rate_result}
  \frac1\tau=\frac{3}{5120\pi^3}
             \frac{\Lambda^2\varepsilon^8}{\hbar^5 m^* v^{10}}
\end{equation}
for the decay rate of a fermionic quasiparticle with energy
$\varepsilon=vk_1$.  The strong suppression of the quasiparticle decay
at low energies, $\tau^{-1}\propto \varepsilon^8$, is our main result.

Similar to the decay rate for weakly interacting fermions, the eighth
power of energy is a combined effect of the weak scattering amplitude,
$|\mathcal A|^2\propto \varepsilon^4$, and small phase space volume
for three-particle scattering, $\nu\propto \varepsilon^4$
\cite{khodas2007fermi, pereira_spectral_2009,
  ristivojevic_relaxation_2013}.  The former result is a direct
consequence of the Fermi statistics of the quasiparticles, whereas the
latter is limited to systems where the quadratic correction to the
spectrum (\ref{eq:varepsilon}) is not forbidden by symmetry.  For
example, the result (\ref{eq:decay_rate_result}) does not apply to
interacting fermions on a tight-binding chain at half filling.  Apart
from this limitation, our result (\ref{eq:decay_rate_result}) is
rather generic.  Most importantly, it is not limited to systems of
weakly interacting fermions.  In particular, it applies at strong
interactions, the so-called Wigner crystal regime, when the
Luttinger-liquid parameter $K\ll1$.  One should keep in mind, however,
that in this case the description of the system in terms of weakly
interacting fermionic excitations is expected to be valid only at
particularly small energies $\varepsilon\ll \sqrt K\,vp_F$
\cite{lin_thermalization_2013}.

The prefactor in our result (\ref{eq:decay_rate_result}) for the decay
rate is expressed in terms of the parameter $\Lambda$ introduced in
Eq.~(\ref{eq:matrix_element_ansatz}).  A microscopic theory for it can
be developed in the limit of weak interactions \cite{khodas2007fermi,
  ristivojevic_relaxation_2013}.  At arbitrary interaction strength,
an analytic microscopic treatment is possible only for integrable
models, in which excitations are expected to have infinite lifetimes,
i.e., $\Lambda=0$.  On the other hand, it is possible to obtain an
expression for $\Lambda$ in terms of the parameters $v$, $m^*$, and
$\lambda$ of the excitation spectrum (\ref{eq:varepsilon}) and their
dependences on the density $n$ and momentum per particle $\kappa$ of
the liquid.

The most straightforward approach involves identifying the possible
perturbations to the Hamiltonian (\ref{eq:H_0}) up to the scaling
dimension 5 and performing the evaluation of the $T$-matrix up to
second order in such perturbations.  This is a laborious procedure
that will be discussed elsewhere.  Here we pursue an alternative
approach based on the idea that a quasiparticle can often be
treated as a mobile impurity in the Luttinger liquid
\cite{ogawa1992fermi, neto1996dynamics, khodas2007fermi,
  imambekov2008exact, pereira_spectral_2009, schecter_dynamics_2011,
  matveev_scattering_2012}.

Luttinger liquid theory applies only to low-energy properties of the
system.  Thus its Hamiltonian (\ref{eq:Luttinger_Hamiltonian})
accounts only for the excitations with energies below certain
bandwidth $D$.  The exact value of $D$ is usually not important, as
long as it is small compared to the characteristic energy scales of
the problem, such as the Fermi energy.  In our discussion so far the
quasiparticles were constructed out of bosonic excitations of the
Luttinger liquid, i.e., the quasiparticle energy $\varepsilon \ll D$.
Alternatively, one can choose $D\ll\varepsilon$, in which case the
quasiparticle is not part of the Luttinger liquid and should be
treated as a mobile impurity.  Let us consider a special case of the
three-particle scattering process (\ref{eq:matrix_element}) for which
$k_2'=-Q$ and $k_2=-Q-\delta Q$, where $Q\gg D/v$, and all the
remaining momenta are such that $|k_1|, |k_1'|, |p|, |p'|\ll D/v$.
According to Eq. (\ref{eq:matrix_element_ansatz}), to leading order in
small $\delta Q$, the scattering matrix element for this process is
given by
\begin{equation}
  \label{eq:low_momentum_scattering}
  \mathcal A = \frac{\Lambda}{L^2}Q^2\,
               \delta_{q-\tilde q, \delta Q},
\end{equation}
where $q=k_1-k_1'$ and $\tilde q=p'-p$ are the momenta of the
particle-hole pairs collapsed near the right Fermi point and created
near the left one, respectively.  

Since the particle-hole pairs correspond to bosons of the standard
Luttinger liquid theory (\ref{eq:Luttinger_Hamiltonian}), one can
think of this process as scattering of a hole from state $Q$ to
$Q+\delta Q$ while absorbing a boson $q$ and emitting a boson $\tilde
q$.  Such processes were studied in
Ref.~\onlinecite{matveev_scattering_2012}.  The expression for the
respective scattering amplitude can be brought to the form
\cite{matveev_equilibration_2013}
\begin{equation}
  \label{eq:t-total}
    t_{q,\tilde q} = -\frac{\sqrt{|q\tilde q|}}{2\pi\hbar L}\,Y_{Q},
\end{equation}
with
\begin{eqnarray}
  \label{eq:Y}
      &&Y_Q=
      \partial^2_{LR}\varepsilon_Q
      -\frac{1}{m_Q^*}
      \frac{\partial_L\varepsilon_{Q}}{v+v_Q}
      \frac{\partial_R\varepsilon_{Q}}{v-v_Q}
     +\partial_L v_Q
      \frac{\partial_R\varepsilon_{Q}}{v-v_Q}
\nonumber\\
   &&\hspace{0.5em} -\partial_R v_Q
      \frac{\partial_L\varepsilon_{Q}}{v+v_Q}
      +\frac{v\partial_n K}{\sqrt K}
      \bigg(
       \frac{\partial_R\varepsilon_Q}{v-v_Q}
       +\frac{\partial_L\varepsilon_Q}{v+v_Q}
      \bigg).
\end{eqnarray}
The quasiparticle velocity and mass here depend on momentum,
$v_Q=\varepsilon_Q'$ and $1/m_Q^*=\varepsilon_Q''$, and the following
shorthand notation is used:
\begin{subequations}
\begin{eqnarray}
  \label{eq:derivatives}\hspace{-2em}
   \partial_R &=& \sqrt{K}\partial_n
             +\frac{\pi\hbar}{\sqrt K}\partial_\kappa,
\\
   \partial_L &=& \sqrt{K}\partial_n
             -\frac{\pi\hbar}{\sqrt K}\partial_\kappa,
\\
  \partial^2_{LR} &=& K\partial_n^2-\frac{(\pi\hbar)^2}{K}\partial_\kappa^2.
\end{eqnarray}
\end{subequations}
Because scattering of the hole by two bosons in the Luttinger liquid
is a special case of the three-fermion scattering event, we can use
the above result to determine $\Lambda$.  To this end, we substitute
the $T$-matrix $t_{q,\tilde q}\,a_{-Q}^\dagger a_{-Q-\delta Q}b_{\tilde
  q}^\dagger b_{q}^{}$ into Eq.~(\ref{eq:matrix_element}) and use
Eq.~(\ref{eq:bosonization}) for the boson operators.  The resulting
scattering amplitude has the form (\ref{eq:low_momentum_scattering})
with $\Lambda=Y_Q/Q^2$.  As expected, at $Q\to0$ the latter expression
has a finite limit
\begin{eqnarray}
  \label{eq:Lambda}
  \Lambda&=&
          \frac12\left(\partial^2_{LR}\frac{1}{m^*}-2\pi\partial_L\lambda\right)
          -\frac{\partial_L v}{4v}\partial_L\frac{1}{m^*}
          +\frac{(\partial_L v)^2}{4m^*v^2}
\nonumber\\ &&
          -\left(\frac{\partial_L v}{4v}
          +\frac{m^*}{2}\partial_L\frac{1}{m^*}\right)
           \left(\partial_{R}\frac{1}{m^*}-2\pi\lambda\right).
\end{eqnarray}
It is worth noting that the absence of the inversion symmetry in the
above expression is caused by our choice of the quasiparticle on the
right-moving branch.

Expression (\ref{eq:Lambda}) relates $\Lambda$ to the parameters
$v$, $m^*$, and $\lambda$ of the quasiparticle spectrum
(\ref{eq:varepsilon}) and their dependences on the particle density
$n$ and momentum per particle $\kappa$ of the liquid.  In combination
with Eq.~(\ref{eq:decay_rate_result}) it gives a complete expression
for the decay rate of fermionic quasiparticles in a spinless quantum
liquid.  Our result is valid at any strength of the interactions
between the physical particles constituting the liquid.  Although our
discussion focused on liquids of spinless fermions, the results are
also applicable to one-dimensional systems of interacting bosons.

Relaxation of excitations in a one-dimensional system with spins was
recently observed in experiment with quantum wires
\cite{barak_interacting_2010}.  To test our results, the spins may be
polarized by a sufficiently strong in-plane magnetic field.  More
generally, the decay of quasiparticles may be observed as the
broadening of sharp features in the density structure factor or
spectral functions.  Both types of measurements can, in principle, be
performed in experiments with two parallel quantum wires.  The density
structure factor controls the Coulomb drag in such devices
\cite{pustilnik_coulomb_2003}, whereas the spectral functions can be
measured in experiments with momentum-resolved tunneling between the
wires \cite{auslaender_spin-charge_2005}.  The same experiments would
also measure the excitation spectrum, enabling a quantitative test of
our results (\ref{eq:decay_rate_result}) and (\ref{eq:Lambda}).

The authors are grateful to L. I. Glazman, M. Pustilnik, and
Z. Ristivojevic for helpful discussions.  This work was supported by
UChicago Argonne, LLC, under contract No.~DE-AC02-06CH11357, by JSPS
KAKENHI Grant No.~24540338, and by the RIKEN iTHES Project.  This work
was supported in part by the National Science Foundation under Grant
No.~PHYS-1066293 and the hospitality of the Aspen Center for Physics.

\end{document}